\documentclass[aps,prb,twocolumn,showpacs,a4paper,superscriptaddress,showkeys]{revtex4}

\usepackage{dcolumn,amsmath,xspace}

\usepackage{graphicx}
\usepackage{dcolumn}
\usepackage{amsmath}
\usepackage{bm}
\usepackage{ulem}

\usepackage{color}

\usepackage{units}

\usepackage{psfrag}

\newcommand{\be}{\begin{equation}}
\newcommand{\ee}{\end{equation}}
\newcommand{\ben}{\begin{eqnarray}}
\newcommand{\een}{\end{eqnarray}}
\newcommand{\beq}{\begin{equation}}
\newcommand{\eeq}{\end{equation}}
\newcommand{\B}{\mathrm{B}}
\newcommand{\NB}{\mathrm{NB}}

\begin{document}

\title{ 
Electronic properties of asymmetrically doped twisted graphene bilayers}

\author{Guy Trambly de Laissardi\`ere}
\email{guy.trambly@u-cergy.fr}

\affiliation{
Laboratoire de Physique Th\'eorique et Mod\'elisation, CNRS, 
Universit\'e de Cergy-Pontoise,
F-95302 Cergy-Pontoise, France.
}

\author{Omid Faizy Namarvar}
\email{omid.faizy@cemes.fr}
\affiliation{CEMES CNRS, 29 rue Jeanne Marvig, F-31055 Toulouse, France}

\author{Didier Mayou}
\email{didier.mayou@neel.cnrs.fr}

\author{Laurence Magaud}
\email{laurence.magaud@neel.cnrs.fr}

\affiliation{Univ. Grenoble Alpes, Institut Neel, F-38042 Grenoble, France}

\affiliation{CNRS, Institut Neel, F-38042 Grenoble, France}

\date{\today}

\begin{abstract}

Rotated graphene bilayers form an exotic class of nanomaterials with fascinating electronic properties governed by the rotation angle  $\theta$. For large rotation angles, the electron eigenstates are restricted to one layer and the bilayer behaves like two decoupled graphene layer. At intermediate angles, Dirac cones are preserved but with a lower velocity and van Hove singularities are induced at energies where the two Dirac cones intersect. At very small angles, eigenstates become localized in peculiar moiré zones. We analyse here the effect of an asymmetric doping for a series of commensurate rotated bilayers on the basis of tight binding calculations of their band dispersions, density of states, participation ratio and diffusive properties. While a small doping level preserves the $\theta$ dependence of the rotated bilayer electronic structure, larger doping induces a further reduction of the band velocity in the same way of to a further reduction of the rotation angle.

\end{abstract}


\maketitle


\section{Introduction}

What remains really surprising with graphene is that all these outstanding electronic and mechanical properties come from a system that is one atomic layer thick.\cite{Wallace47,Berger06,Castro09_RevModPhys}  
Few layer graphene and more precisely bilayers also present fascinating properties. It has been known for years that in this case, stacking plays a crucial role. While AA bilayers --all C atoms are in the same position in the two layers-- result in two Dirac cones shifted in energy, AB stacking --as in graphite-- breaks the atom A / atom B symmetry and leads to a quadratic 
dispersion.\cite{Latil06,MacdoABC,Varchon_prb08,Ohta06,Brihuega08}  
Here we focus on exotic bilayers that present neither AA nor AB stacking but with a relative rotation of the two layers. 

Different approaches are used nowadays to obtain graphene: mechanical peeling of graphite, annealing of SiC, CVD on metals. These three approaches also give multilayers with, in some cases, a rotation between successive layers. Indeed rotated bilayers have been obtained on graphite but also on Ni and on the C-face of SiC. 
A rotation between two layers creates a (pseudo) periodicity that appears as a moir\'e pattern on STM 
images.\cite{Hass08,Emtsev08,Varchon08,Sprinkle09}
All the theoretical works \cite{Latil07, Dossantos07, Shallcross08, Shallcross10, Bistritzer10,Bistritzer11, Bistritzer11_PNAS, Mele10, Mele11, Suarez10, Trambly10, Trambly12, LopesDosSantos12, Suarez13,Omid14, Uchida14,Landgrafet13,Sboychakov15} now agree on the fact that two graphene layers stacked with a rotation between them show exotic electronic properties that are angle-dependant. The AA and AB stacking are the two extreme cases, they correspond to rotations of 0$^o$ and 60$^o$. The bilayer behavior is symmetric with respect to a rotation angle equal to 30$^o$. At large angles (close to 30$^o$), the two layers are decoupled and behave like independent graphene planes. At smaller angles, graphene Dirac cones are conserved but the velocity is renormalized (reduced). 
Van Hove singularities (vHs) are found at energies where the Dirac cones from the two layers intersect.\cite{Li10,Andrei_Ni11,Brihuega12,Cherkez15} 
Eventually for small angles the two vHs merge at the Dirac energy and give a sharp peak in the density of states (DOS). The corresponding states are localized in a region of the supercell where stacking is close to AA.\cite{Trambly10,Trambly12}


Here we check the robustness of the theoretical predictions with respect to doping which can be an important perturbation. Indeed, bilayers often show an asymmetric doping --one layer more doped than the other one-- either the doping is made on purpose 
if a potential bias is applied between the layers, or it  
results from charge transfer with a substrate. 
In a tight-binding (TB) scheme, an asymmetric doping is a shift in electrochemical potential between the two layers. 
An asymmetric doping opens a gap in the band structure of an AB bilayer. We will show that it is not the case for rotated bilayers and that for not too small angles and reasonable doping, linear dispersion, velocity renormalization and vHs remain.  The main effect of doping is to shift one Dirac cone with respect to the other one by an energy that varies with the doping rate and the rotation angle. Localization of the states either on one layer (large angle, decoupled layers) or on both but in AA regions (small angles) is not drastically changed by doping.  
The complex electronic structure of graphene bilayers is a consequence of the local geometry of the system. A parallel can be drawn with quasicrystals where the quasicrystals specific properties develop when the size of the approximant cell increases.\cite{Trambly06,Trambly14_QC,Trambly14}  In the same way, here specific properties arise when the commensurate cell size increases  and the AA and AB regions are better defined. The parallel is obvious when one look at transport properties and the importance of the non-Boltzmann part either for neutral or doped bilayers.

The numerical method and atomic structures of rotated bilayer are detailed in Sec. \ref{SecMethds} and in the appendix, then the effect of doping on the band structure (Sec. \ref{SecBands}), average velocity (section \ref{SecAverageVelocity}) and the density of states (Sec. \ref{SecDOS}) is discussed. 
The participation ratios are convenient quantity to characterize the states repartition as a function of the energy. It is shown for neutral and doped graphene in Sec. \ref{Secparticipation_ratio}.  
Finally specific quantum diffusion due to confined states in doped and undoped twisted graphene bilayers are presented Sec. \ref{SecTransportBilayer}. For comparison quantum diffusion in graphene is presented in the appendix.

\section{Numerical methods and atomic structure}
\label{SecMethds}
\begin{figure}
\includegraphics[width=8cm]{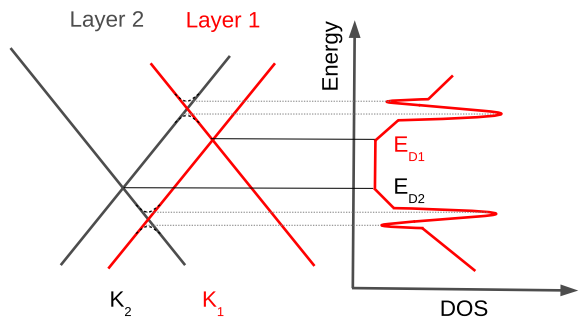}

\caption{Dirac cones and total DOS of a doped rotated bilayer, schematic diagram.
}
\label{Fig_vHs_dop}
\end{figure}

\begin{table}
\caption{\label{Tab_bilayer} 
Studied $(n,m)$ bilayer structures. $N$ is the number of atoms, $\theta$ the rotation angle and  $V_{x} / V_{\rm m}$ the undoped rotated bilayer velocity at K point along $x$-direction divided by the monolayer velocity.\cite{Trambly12}
}
\begin{tabular}{lrrr}
\hline \hline
{($n,m$)}~~~  &  ~~$\theta$ ($^o$)   &  ~~~~$N$  &  ~$V_{x} / V_{\rm m}$    \\ \hline
(1,3)    &  32.20     &   52    &   0.99    \\
(5,9)    &  18.73     &   604   &  0.99     \\
(2,3)    &  13.17     &   76    &  0.96     \\
(3,4)    &   9.43    &  148     &  0.95     \\
(6,7)    &  5.08     &  508     &  0.83    \\
(8,9)    &  3.89     &  868     &  0.74     \\
(12,13)    &  2.65     &  1876     &  0.48     \\
(15,16)    &  2.13     &  2884     &  0.35     \\
(25,26)    &  1.30     &  7804     &  0.02     \\
(33,34)    &  0.99     &  13468     & 0.01      \\
\hline \hline
\end{tabular}
\end{table}

Tackling small rotation angles --smaller than 4$^o$-- means handling very large cells that can involve a huge number of atoms i.e. more than 
10000 (table \ref{Tab_bilayer}). 
We use a tight binding (TB) scheme developped\cite{Trambly10,Trambly12}
for $p_z$ orbitals since we are interested in what happens at energies within $\pm 2$\ eV of $E_{\rm D}$, the Dirac point energy whatever the rotation angle is. 
The TB scheme is described in details in Ref.\,\onlinecite{Trambly12}. 
Since the planes are rotated, neighbours are not on top of each other 
(as it is the case in the Bernal AB stacking). Interlayer interactions are then not restricted to 
$pp\sigma$ terms but some  $pp\pi$ terms have also to be introduced. 
The Hamiltonian has the form :
\ben
\hat{H} &=& \sum_{i} \epsilon_i \, |i\rangle \langle i|  ~+~  \sum_{<i,j>} t_{ij} \, |i\rangle \langle j| ,
\label{Eq_hamilt}
\een
where  $|i\rangle$ is the $p_z$ orbital with energy $\epsilon_i$ located at $\vec r_i$, and $\langle i,j\rangle$ is 
the sum on index $i$ and $j$ with $i\ne j$.
The coupling matrix element, $t_{ij}$, 
between two $p_z$ orbitals located at $\vec r_i$ and $\vec r_j$ 
is,\cite{SlaterKoster54}
\ben
t_{ij} ~=~ \langle i|\hat{H}|j\rangle 
~=~ n_c^2 V_{pp\sigma}(r_{ij}) ~+~ (1 - n_c^2) V_{pp\pi}(r_{ij}),
\een
where $n_c$ is the direction cosine, $V_{pp\sigma}$ and $V_{pp\pi}$ the Slater-Koster coupling parameters.
In our scheme,\cite{Trambly12} $V_{pp\sigma}$ and $V_{pp\pi}$ are exponentially decaying function of the distance. 
It is known that the results of the band calculations are sensitive to a particular form of these parameters, and different parametrizations of the Slater-Koster coupling parameters are used in the literature.\cite{Landgrafet13,Sboychakov15}
However many general aspects of the band structure in rotated twisted bilayer are found similarly with different TB parametrizations.
Asymmetric doping is modeled using different on-site energies on the two layers. All orbitals of a layer have the same energy.  
In the following, results are given as a function of the potential bias $\Delta \epsilon$ between the two layers.
$\Delta \epsilon$ is the difference between the on-site energies $\epsilon_1 = \epsilon_0$ on the top and the on-site energies $\epsilon_2 = \epsilon_0 - \Delta \epsilon$ on the bottom layers. The coupling beyond first neighbor induces an asymmetry between states above and below the Dirac energy in each layers. All energies are then given with respect to the Dirac energy $E_{D1}$ of the top undoped layer (top layer) which is set to zero.

The eigenstates obtained by diagonalisation in reciprocal space of the TB Hamiltonian are used to calculate transport characteristic values (velocity, square spreading, diffusivity) as explained in the appendix \ref{SecMethodeQuantumTransp}.
In monolayer graphene (appendix \ref{SecMonoGraphene}), transport properties are well described by the usual semi-classical Boltzmann approach (excepted at Dirac energy), 
but in twisted bilayer with small rotation angle $\theta$, very unusual effects occur 
at Dirac energy that are not taken into account by Boltzmann approach (section \ref{SecTransportBilayer}). 
The average densities of states in each layer are presented briefly in section \ref{SecDOS}. They are calculated by recursion method in real space starting from a random phases state.\cite{Roche97} This method gives total DOSs that are similar to the one obtained by diagonalisation in reciprocal space.

Our calculations require periodic boundary conditions. The way a bilayer supercell is built and how it is labeled $(n,m)$ is described in Refs. [\onlinecite{Trambly10,Trambly12}].
We start from an AA bilayer and choose the rotation origin O  at an atomic site. 
A commensurate structure can be defined if the rotation changes a lattice vector 
${\overrightarrow{OB}} ~(m,n)$ to
${\overrightarrow{OB'}}~(n,m)$, where the integers $n$, $m$ are the coordinates with respect to the basis vectors 
$\vec a_1$ $(\sqrt{3}a/2,-a/2)$  
and $\vec a_2$ $(\sqrt{3}a/2,a/2)$, with $a=0.2456$\,nm. 
The rotation angle is then defined by:
\begin{equation}
\cos\theta = \frac{n^2+4nm+m^2}{2(n^2+nm+m^2)},
\label{eq:Comm3}
\end{equation}

and the commensurate cell vectors correspond to:
\begin{equation}
\vec t = {\overrightarrow{OB'}} = n \vec a_1 + m \vec a_2\,,~ \\
\vec{t'} = -m \vec a_1 + (n+m) \vec a_2.
\label{eq:Comm4}
\end{equation}

The commensurate unit cell contains $N = 4(n^2+nm+m^2)$ atoms. It is now well established \cite{Dossantos07,Suarez10,Trambly10,Trambly12,LopesDosSantos12,Suarez13,Omid14} that  
the rotation angle $\theta$ is a good parameter to describe the system but the number of atoms is not, since cells 
of equivalent size can be found for different angles. 
For $\theta$ values less than $\sim 15^o$, twisted bilayer form a moir\'e pattern with (pseudo) period $P$, \cite{Campanera07}
\begin{equation}
P = \frac{a}{2 \sin(\theta/2)}   \simeq \frac{ 1.42 }{ \theta ({\rm deg})} {\rm ~in~ nm}.
\label{EqPeriodeMoire}
\end{equation}

Structures of the bilayers studied in this paper are listed in table \ref{Tab_bilayer}. 

\section{Band dispersions}
\label{SecBands}

\subsection{Large and intermediated angles}

\begin{figure}
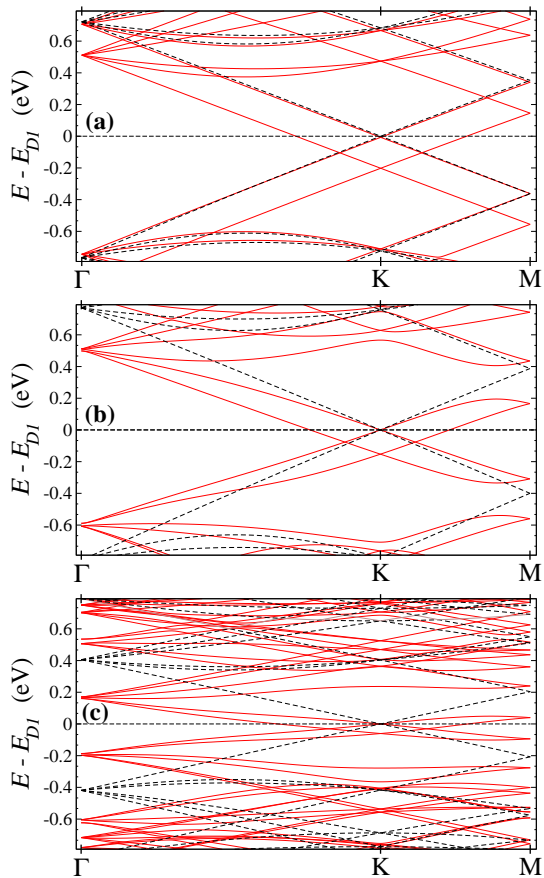


\includegraphics[width=7cm]{FigDec13_Bi59_bnds_DE.2.eps} 

\vskip .1cm
\includegraphics[width=7cm]{FigDec13_Bi67_DE.2_bnds_TB.eps}

\vskip .1cm
\includegraphics[width=7cm]{FigDec13_bi_12-13_DE.2_bnds.eps}

\caption{(color online) Band dispersions for 
doped bilayers with $\Delta \epsilon = 0.2$\,eV.
(a) (5,9) bilayer ($\theta = 18.73^{\rm o}$),
(b) (6,7) bilayer  ($\theta = 5.08^{\rm o}$)
and 
(c) (12,13) bilayer ($\theta = 2.65^{\rm o}$).
Lines (dashed-line) are TB calculations for bilayers (monolayer). 
}
\label{fig_bnds_dopees}
\end{figure}

\begin{figure}
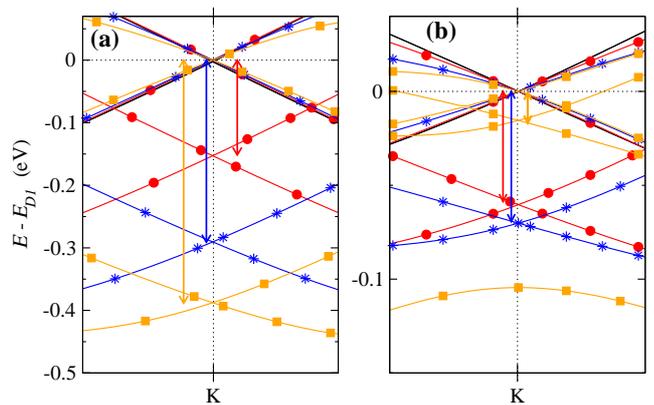


\includegraphics[height=5.3cm]{FigDec13_bi_6-7_DE.2_.4_.6_bnds.eps}
\includegraphics[height=5.3cm]{FigDec13_bi_12-13_DE.2_.4_.6_bnds.eps}

\caption{(color online)  Band dispersions for 
doped bilayers: 
(a) (6,7) bilayer  ($\theta = 5.08^{\rm o}$),
(b) (12,13) bilayer ($\theta = 2.65^{\rm o}$).
The on-site energy difference  $\Delta \epsilon$ 
between $p_z$ orbitals in the two layers
is 
(solid line) $\Delta \epsilon = 0$,
(circle) $\Delta \epsilon = 0.2$\,eV,
(star) $\Delta \epsilon = 0.4$\,eV,
(square) $\Delta \epsilon = 0.6$\,eV.
Arrow show the energy difference $\Delta E_D$ between bands at {\bf K}
(Table \ref{Tab_DE}).
$E(\vec k)$ has been computed for more then 50 $\vec k$ values in the k-scale shown in these figures.
}
\label{fig_bnds_dopees2}
\end{figure}

\begin{table}
\caption{\label{Tab_DE} 
Energy bands splitting $\Delta E_D$ at K point in doped ($n,m$) bilayers. 
$\Delta \epsilon$ is the  on-site energy difference
between $p_z$ orbitals in the two layers.
}
\begin{tabular}{lll}
\hline \hline
{($n,m$)}~~~~~~  &  $\Delta \epsilon$ (eV)~~~~   &  $\Delta E_D $ (eV)   \\ \hline
(5,9)         &  0.2  &  0.20      \\
              &  0.4  &  0.40      \\
              &  0.6  &  0.60      \\
(3,4)         &  0.5  &  0.45  \\
              &  1.0  &  0.87  \\
              &  1.25  & 1.00     \\
(6,7)         &  0.2  &  0.15 \\
              &  0.4  &  0.29      \\
              &  0.6  &  0.39      \\
(12,13)       &  0.2  &  0.06  \\
              &  0.4  &  0.07      \\
              &  0.6  &  0.02      \\
(25,26)  &  0.2  &  0.011      \\
         &  0.4  &  0.004      \\
\hline \hline
\end{tabular}
\end{table}

\begin{figure}

\includegraphics[width=8cm]{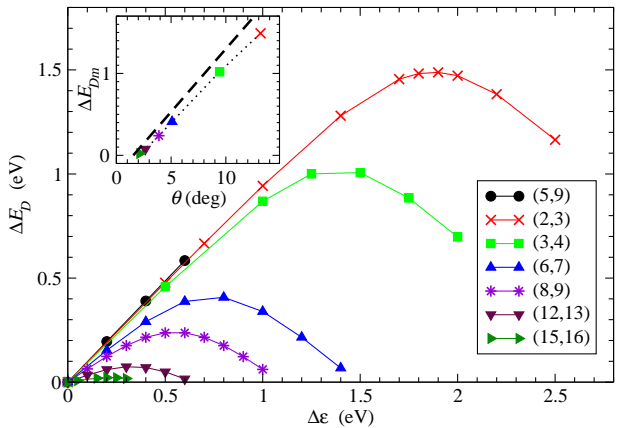}

\caption{(color online) 
$\Delta E_D$ versus $\Delta \epsilon$ in rotated doped bilayers
[Insert: Maximum value, $\Delta E_{Dm}$,  of $\Delta E_D$ versus the $\theta$ angle. Points line is guide for the eyes.
The dashed lines show $\Delta E_{\rm vHs}$ given by equation (\ref{Eq_DEvHs}) 
i.e. here $\Delta E_{\rm vHs} \simeq 0.15\, \theta (Deg) - 0.24$  ].
}
\label{fig_DEfDEpsilon}
\end{figure}
 
\begin{figure}
\includegraphics[width=7cm]{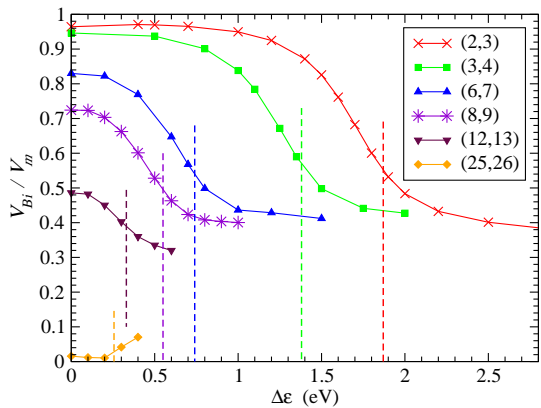}

\caption{(color online) Velocity at Dirac point (slope of the band along KM at K and Dirac energy) 
versus $\Delta \epsilon$ in rotated doped bilayers. The vertical dashed lines show the values $\Delta \epsilon_m$ 
for which $\Delta E_D(\Delta \epsilon)$ is maximum.
}
\label{Figure_VB_atK}
\end{figure}

\begin{figure}
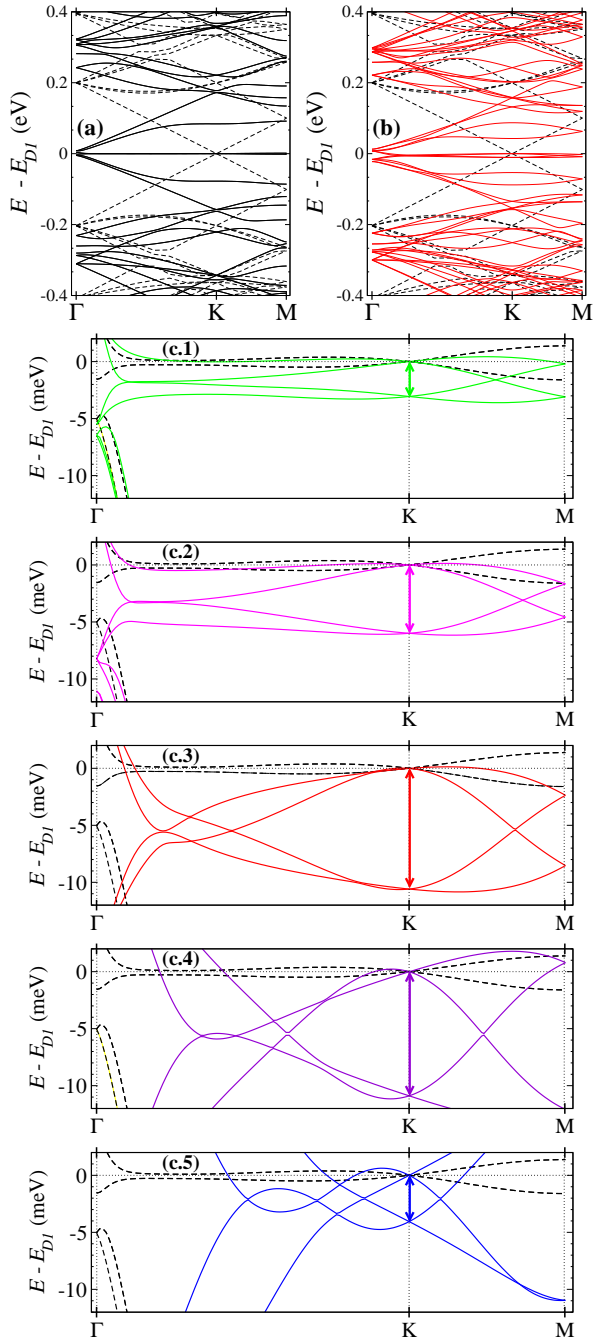


\includegraphics[height=4.2cm]{Fig_bi_25-26_bnds_De0_mars15.eps}\hspace{.1cm}\includegraphics[height=4.2cm]{Fig_bi_25-26_bnds_De.2_mars15.eps}

\vskip .1cm
\includegraphics[width=7.2cm]{Fig_bi_25-26_bnds_aout15_DE.05.eps}
\vskip .1cm
\includegraphics[width=7.2cm]{Fig_bi_25-26_bnds_aout15_DE.1.eps}
\vskip .1cm
\includegraphics[width=7.2cm]{Fig_bi_25-26_bnds_aout15_DE.2.eps}
\vskip .1cm
\includegraphics[width=7.2cm]{Fig_bi_25-26_bnds_aout15_DE.3.eps}
\vskip .1cm
\includegraphics[width=7.2cm]{Fig_bi_25-26_bnds_aout15_DE.4.eps}

\caption{(color online) Band dispersions for (25,26) bilayer ($\theta = 1.30^{\rm o}$):
(a) undoped bilayer ($\Delta \epsilon= 0$),
(b) doped bilayer $\Delta \epsilon= 0.2$\,eV.
(c) Zoom arround energy $E_{D1}$: 
(dashed line) $\Delta \epsilon= 0$, 
(c.1)  $\Delta \epsilon= 0.05$\,eV,
(c.2)  $\Delta \epsilon= 0.1$\,eV,
(c.3)  $\Delta \epsilon= 0.2$\,eV,
(c.4)  $\Delta \epsilon= 0.3$\,eV,
(c.5) $\Delta \epsilon= 0.4$\,eV.
The arrow show the energy difference $\Delta E_D$.
}
\label{Fig_bnds_25-26}
\end{figure}

An asymmetrically doped bilayer with a bernal stacking presents a gap due to the break of all atom A / atom B symmetry. But in twisted bilayer the situation is completely different.\cite{McCann06}
A schematic diagram of the asymetrically doped rotated bilayer is given in figure \ref{Fig_vHs_dop}. It applies to the large and intermediate angle cases which still show two Dirac cones. The small angle limit is more complex because of the important state mixing between the two layers. As a consequence of the asymetric doping, one Dirac cone is shifted and intersection between bands no longer occur at the mid point between K$_1$ and K$_2$, as it was the case in neutral systems. Then the maximum of the band (and then the van Hove singularity at $E_+$ and $E_-$, see Sec. \ref{SecDOS}) is no longer located at point M of the supercell Brillouin zone (Figs. \ref{fig_bnds_dopees} and \ref{fig_bnds_dopees2}).
No gap opens even for large doping ($\Delta \epsilon = 0.6$\,eV and even more). For a given doping, the energy difference, 
$\Delta E_D = E_{D1} - E_{D2}$, between the two Dirac points varies with the rotation angle which results from interplane state mixing. 
These energy differences are shown by arrows on figure \ref{fig_bnds_dopees2} and they are given in table \ref{Tab_DE}.

Considering the two Dirac cones in  reciprocal space (figure \ref{Fig_vHs_dop}),
whatever the values of $ \Delta \epsilon > 0$ and of $\theta$ are, 
TB calculations show that
the bands of the shifted Dirac cone never cross the bands of the non-shifted Dirac cone.
Then $\Delta E_D(\Delta \epsilon)$ has a maximum, $\Delta E_{Dm}$, for every $\theta$ value 
(see the insert of figure \ref{fig_DEfDEpsilon}).
This limit $\Delta E_D$ value, corresponding to $\Delta \epsilon = \Delta \epsilon_m$, is always found when one branch of the shifted Dirac cone approaches the parallel branch of the second cone 
($\Delta E_D(\Delta \epsilon_m) = \Delta E_{Dm}$).
If doping is small enough, $\Delta \epsilon < \Delta \epsilon_m$,   the energy difference $\Delta E_D$ between the Dirac cones increases with the on-site energy differences $\Delta \epsilon$. The increase factor is smaller for smaller angles. 
If doping is larger, $\Delta \epsilon > \Delta \epsilon_m$,  $\Delta E_D$ decreases when doping increases  (figure \ref{fig_DEfDEpsilon}).
For each angle $\theta$, the maximum value can be understood in a simple scheme as follows. 
For small doping --small $\Delta \epsilon$-- it is obvious that 
\be
\Delta E_{D} ~\le~ \Delta E_{Dm} ~<~ \Delta E_{\rm vHs} ~=~ E_+ - E_- . 
\label{Eq_ConditionEDmax}
\ee
This condition is satisfied for any dopping (at least for dopping that preserves the existence of Dirac cone). 
From continuum model,\cite{Dossantos07,Li10,LopesDosSantos12,Suarez13} experimental measurements\cite{Brihuega12,Cherkez15} 
and our calculations (section \ref{SecDOS}), 
the energy of van Hove singularity is: 
\be
\Delta E_{\rm vHs} ~=~ \hbar v_F {\rm K_1 K_2}  - 2t_{\theta} ~=~ 2 \hbar v_F \Gamma{\rm K} \sin(\theta / 2) - 2t_{\theta}
\label{Eq_DEvHs}
\ee
for angles larger than $\sim 2^{\rm o}$. 
Where $v_F$ is the Fermi velocity for monolayer graphene, $\Gamma{\rm K}=1.703 \, {\rm \AA}^{-1}$ is the wave vector of Dirac point in monolayer graphene, and $t_{\theta}$ is the modulus of the amplitude of the main Fourier compoments of the interlayer potential, $t_{\theta} \simeq 0.12$\,eV.\cite{Trambly12,Omid14} The calculated maximum value of $\Delta E_D$ is drawn in the insert of figure \ref{fig_DEfDEpsilon}, showing that condition (\ref{Eq_ConditionEDmax}) is satisfied. 
It is interresting to remark also that for $(n,m)$ bilayer, such as $|m-n|>1$ for instance (5,9) figure \ref{fig_DEfDEpsilon}, 
when $\Delta \epsilon$ increases a mixing of bands could occur before the maximum value of $\Delta E$ estimated by the condition (\ref{Eq_ConditionEDmax}). 

For doping small enough, $\Delta \epsilon \ll \Delta \epsilon_m$,
slopes of the band dispersions $E(k)$ at Dirac point 
are not modified (figures \ref{fig_bnds_dopees}(a), \ref{fig_bnds_dopees}(b) and \ref{fig_bnds_dopees2}(a)).
But for larger $\Delta \epsilon$ values, this slopes decreases as  $\Delta \epsilon$ increases, which results in a strong reduction of the intra-band velocity at Dirac points. 
Figure \ref{Figure_VB_atK}, shows this renormalization for different $\theta$ values.
Above this limit (figures \ref{fig_bnds_dopees}(c) and \ref{fig_bnds_dopees2}(b)), 
$\Delta \epsilon > \Delta \epsilon_m$,
bands become flatter and intra-band velocity reaches a limit value $\sim 0.4 V_{mono}$ (figure \ref{Figure_VB_atK}).
Therefore for large rotation angles and physicaly reasonnable doping ($\Delta \epsilon \ll \Delta \epsilon_m$), the velocity renormalisation \cite{Dossantos07,Suarez10,Bistritzer11, Bistritzer11_PNAS,Trambly10,Trambly12,LopesDosSantos12,Suarez13,Omid14,Uchida14} is not modified;
but for intermediate rotation angles, actual doping can lead to strong velocity renormalisation. 

\subsection{Very small angles}

The case of very small angles, typically for $\theta < \, \sim2^o$, is illustrated on figure \ref{Figure_VB_atK} and figure \ref{Fig_bnds_25-26} for different doping values in (25,26) bilayer. 
The two Dirac cones at $E_{D1}$ and $E_{D2}$ are still present and the maximum value of $\Delta E_D$ is obtained for 
$\Delta \epsilon_m \simeq 0.25$\,eV (figure \ref{Fig_bnds_25-26}), 
but the behavior of intra-band velocity at K point versus $\Delta \epsilon$ differs from that for intermediate and large angles (figure \ref{Figure_VB_atK}), showing that a new regim is obtained.
Bands with energy $E$ arround Dirac energies are very flat and states at these energies are not only those of Dirac cones at K points. Therefore the velocity of electrons at these energies is the average of velocity of all states $\vec k$ at energy $E(\vec k)=E$ as discussed is next section.

\section{Average intra-band velocity}
\label{SecAverageVelocity}

\begin{figure}
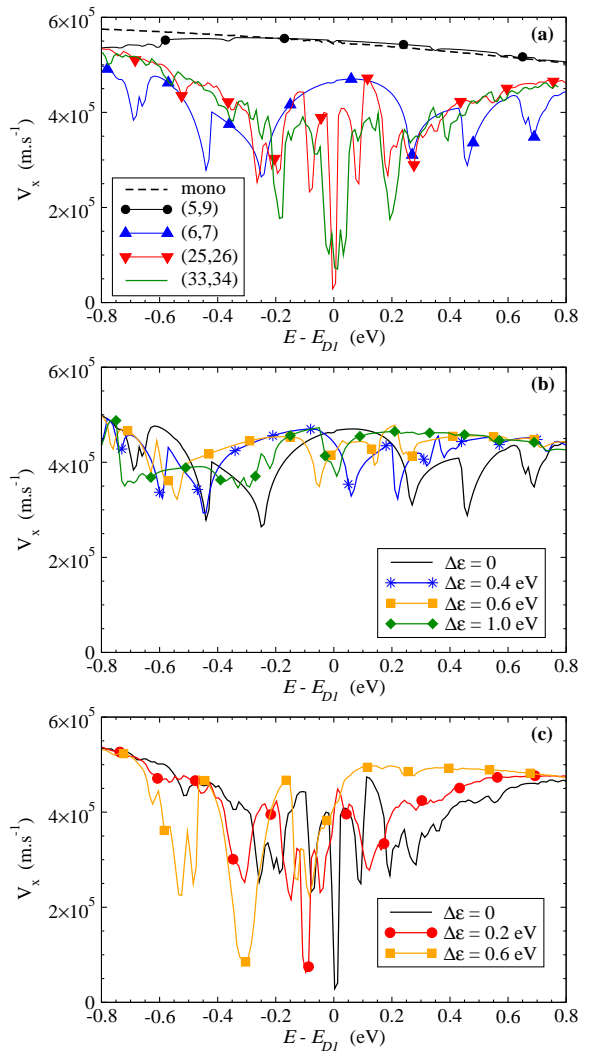

\includegraphics[width=7.5cm]{Fig_VB_AtD_5-9_6-7_25-26_33-34_4dec14.eps}

\includegraphics[width=7.5cm]{Fig_VB_AtD_6-7_DEps.eps}

\includegraphics[width=7.5cm]{Fig_VB_AtD_25-26_Deps_4dec14.eps}

\caption{(color online) Average velocity $V_x$ along X-axis, 
($V = \sqrt{2}V_x$) versus energy $E$. 
(a)
(Dashed line) graphene and undoped ($n,m$)  bilayers ($\Delta \epsilon = 0$).
(b) (6,7) bilayer and (c) (25,26) bilayer:
(solid line) undoped ($\Delta \epsilon= 0$), 
(circle) $\Delta \epsilon =0.2$\,eV, 
(star) $\Delta \epsilon =0.4$\,eV, 
and (square)  $\Delta \epsilon =0.6$\,eV.
}
\label{Figure_VB_diag}
\end{figure}

\begin{figure}
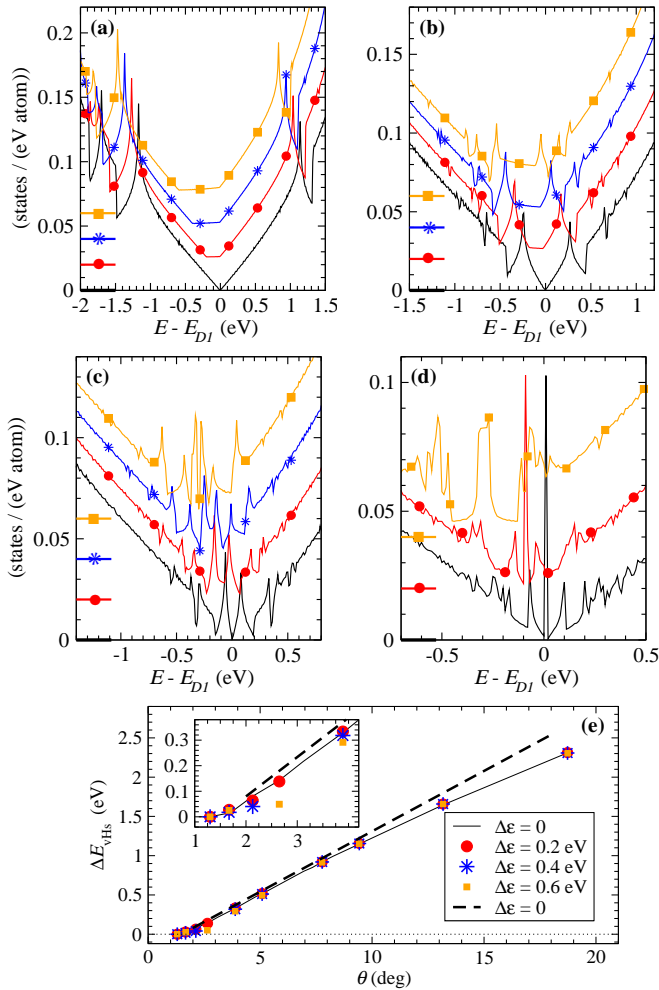


\includegraphics[height=4.5cm]{FigDec13_Bi59_DOS_DE_mars15.eps}
~~\includegraphics[height=4.5cm]{FigDec13_Bi67_DOS_DE_mars15.eps}

\vskip .1cm
\includegraphics[height=4.5cm]{FigDec13_bi12-13_DOS_DE_mars15.eps}
~~~\includegraphics[height=4.5cm]{FigDec14_bi_25-26_DOS_DE_mars15.eps}

\vskip .1cm
~~~~~\includegraphics[width=7cm]{FigBidec15_EVanHove_dope.eps}

\caption{(color online) Total density of states (DOS) in 
(a) (5,9), 
(b) (6,7), 
(c) (12,13)
and (d) (25,26) bilayers.
The on-site energy difference  $\Delta \epsilon$ 
between $p_z$ orbitals in the two layers
is 
(solid line) $\Delta \epsilon = 0$,
(circle) $\Delta \epsilon = 0.2$\,eV,
(star) $\Delta \epsilon = 0.4$\,eV,
(square) $\Delta \epsilon = 0.6$\,eV.
The DOS curves have been shifted vertically for clarity (the origin of the DOS
for each curve is indicated by the horizontal line on the lower left corner).
(e) Energy difference between the energies $\Delta E_{\rm vHs}$ of vHs, $\Delta E_{\rm vHs} = E_+ - E_-$, versus rotation angle for different doping values.
The fine solid line is guide for the eye.
Dashed line shown $\Delta E_{\rm vHs}$ given by equation (\ref{Eq_DEvHs}) 
i.e. here $\Delta E_{\rm vHs} ({\rm eV}) \simeq 0.15\, \theta ({\rm Deg}) - 0.24$.
}
\label{Fig_DOS_DE}
\end{figure}

\begin{figure}
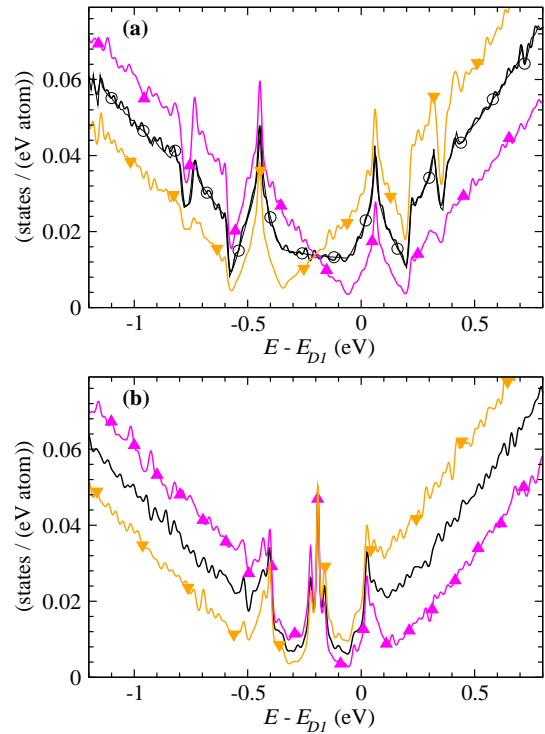


\includegraphics[width=7cm]{Fig_DOSplan_Bi6-7_DE.4_v3.eps}

\vskip .1cm
\includegraphics[width=7cm]{Fig_DOSplan_Bi25-26_DE.4_v3.eps}

\caption{(color online) 
Density of states (DOS) calculated by recursion in doped (a) (6,7), 
(b) (25,26) bilayers with $\Delta \epsilon = 0.4$\,eV:
(solid line) Total DOS,
(triangle up) average DOS in layer 1 (undoped layer), 
(triangle down) average DOS in layer 2 (doped layer). 
Fig (a) the total DOS in (6,7) bilayer, calculated by diagonalisation in reciprocal lattice, is also shown with fine line with empty circle. It is very close with total DOS calculated by recursion method.
}
\label{Fig_DOSlocale}
\end{figure}

The average intra-band velocity (Bloch-Boltzamnn velocity) is calculated numerically from the velocity operator $\hat{V}_x$ along the $x$-direction and equation (\ref{Eq_calcul_VB}) as explained in the appendix \ref{SecMethodeQuantumTransp}.   
It is shown figure \ref{Figure_VB_diag} for several $(n,m)$  bilayers. 
As expected, for large rotated angles and small doping, this method gives velocity values that are similar to the ones calculated directly from the slope of bands $E(\vec k)$ of Dirac cone (intra-band velocity at K shown figure \ref{Figure_VB_atK}).
For intermediate angles (figure \ref{Figure_VB_diag}(b)), the  effect of the renormalisation of the intra-band velocity at K points is seen, but this effect is small because other bands contribute also to the average velocity at same energies.
For very small angles (figure \ref{Figure_VB_diag}(c)), a very small average velocity is obtained at Dirac energy (confined states), 
with velocity similar to the intra-band value at K points.
This renormalization effect remains strong for doped bilayers but energies of localized states (small velocity) are shifted. This strong reduction of the intra-band velocity have consequences on electronic transport properties as shown in section \ref{SecTransportBilayer}.

\section{Density of states}
\label{SecDOS}

The shift of one Dirac cone in doped bilayers induces a modification in the DOS as schematically shown in figure \ref{Fig_vHs_dop}. The van Hove singularities (vHs) are not at the M point of the supercell brillouin zone but fall somewhere on the K--M line. Furthermore, the DOS is constant in between the two Dirac cones --as it is for this energy range in a AA bilayer--. These two caracteristics are found on the DOS of bilayers with large and intermediate angles figures \ref{Fig_DOS_DE}(a,b). 
For realistic doping and not too small rotation angles, the variations 
of the vHs energies difference $\Delta E_{\rm vHs}$, $\Delta E_{\rm vHs} = E_+ - E_-$,  with the rotation angle are very similar to those of undoped bilayers (figure \ref{Fig_DOS_DE}(e)), as recently found from Scanning Tunneling Spectroscopy by V. Cherkez et al.\cite{Cherkez15}
For very small angles, the localization in AA zone of the moir\'e is still present and the sharp
peak in the DOS is shifted (figures \ref{Fig_DOS_DE}(d) and \ref{Figure_VB_diag}(c)). 

DOSs in each layer of doped bilayers are presented figure \ref{Fig_DOSlocale} for intermediate and small rotation angles. As expected the global shape of DOS in doped layer is shifted in energy by the doping. In any cases the peaks of vHs or the peak of localization around Dirac energy are clearly seen in the two layer DOSs at the same energies. 
This suggests that corresponding states are spread in the two layers as shown in the next section. 

\section{Participation ratio}
\label{Secparticipation_ratio}

\begin{figure}
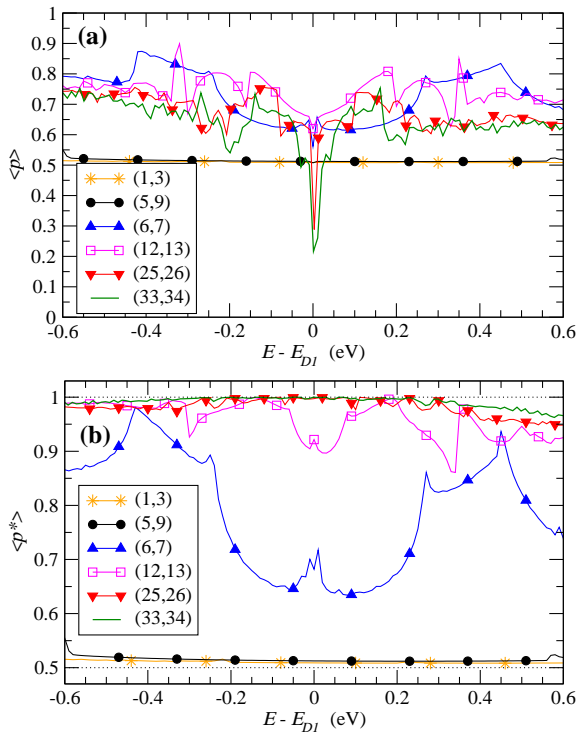


\includegraphics[width=7.5cm]{FigDec14_part_ratio.eps}

\vskip .1cm
\includegraphics[width=7.5cm]{FigDec14_part_ratio_plan.eps}

\caption{(color online) TB (a) average participation ratio 
and (b) average layer participation ratio  in ($n$,$m$) undoped bilayers ($\Delta \epsilon = 0$).
}
\label{Fig_participaionRatio}
\end{figure}

\begin{figure}
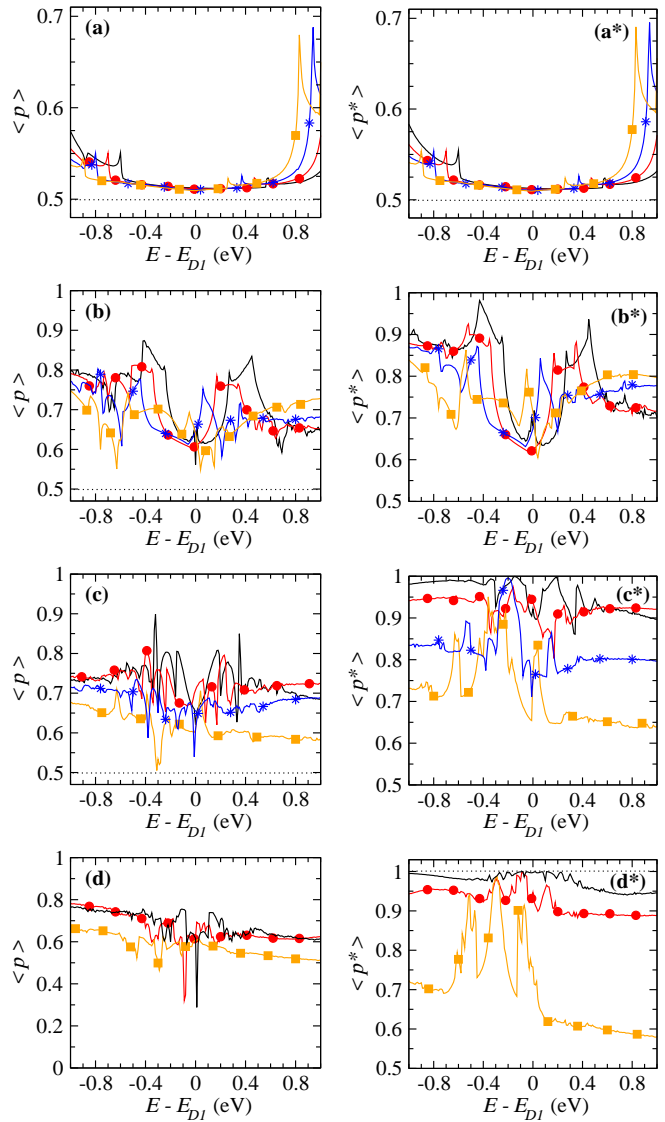

\includegraphics[width=4.1cm]{FigDec13_Bi59_PartRatio_DE.eps}
~~\includegraphics[width=4.1cm]{FigDec13_Bi59_PartRatioPlan_DE.eps}

\vskip .1cm
\includegraphics[width=4.1cm]{FigDec13_bi_6-7_PartRatio_DE.eps}
~~\includegraphics[width=4.1cm]{FigDec13_bi_6-7_PartRatioPlan_DE.eps}

\vskip .1cm
\includegraphics[width=4.1cm]{FigDec13_bi12-13_PartRatio_DE.eps}
~~\includegraphics[width=4.1cm]{FigDec13_bi12-13_PartRatioPlan_DE.eps}

\vskip .1cm
\includegraphics[width=4.1cm]{FigDec13_bi_25-26_PartRatio.eps}
~~\includegraphics[width=4.1cm]{FigDec13_bi_25-26_PartRatioPlan.eps}

\caption{(color online) TB average participation ratio $<${$p$}$>$
and average layer participation ratio $<${$p$}*$>$ in 
(a) (5,6), 
(b) (6,7), 
(c) (12,13)
and (d) (25,26) bilayers.
The on-site energy difference  $\Delta \epsilon$ 
between $p_z$ orbitals in the two layers
is 
(black line) $\Delta \epsilon = 0$,
(circle) $\Delta \epsilon = 0.2$\,eV,
(star) $\Delta \epsilon = 0.4$\,eV,
(square) $\Delta \epsilon = 0.6$\,eV.
}
\label{Fig_participaionRatio_dope}
\end{figure}

To analyse the nature of the eigenstates in the bilayers and search for a possible doping effect, we 
compute the participation ratio of each TB eigenstate $| \psi \rangle$
defined by
\be
p(\psi) = \frac{1}{N \, \sum_i |\langle i | \psi \rangle|^{4} }\,,
\ee
where $| i\rangle$ are the $p_z$ orbitals on  atoms $i$ and $N$ is the number of 
atoms in a unit cell.
For a completely delocalized eigenstate, $p$ is equal to $1$
as in graphene. If the state is restricted to one graphene layer, $p$ is equal to 0.5 and a state localized on 1 atom have the smallest 
$p$ value: $p=1/N$.  
The average participation ratio $\langle p\rangle$ as a function of the energy $E$ is presented on 
figure \ref{Fig_participaionRatio}(a) for a neutral bilayers and figure \ref{Fig_participaionRatio_dope}(a-d) for a doped ones. 

The participation ratios for neutral systems clearly illustrate the three regimes of the electronic structure of twisted bilayers as a function of the rotation angle through the behavior of the eigenstates.

For large angles $\theta$ --bilayers $(1,3)$ ($\theta=32.20^o$) and $(5,9)$ ($\theta=18.73^o$) in figure \ref{Fig_participaionRatio}(a)-- 
$\langle p\rangle$ is  equal to 0.5 which means that the eigenstate is delocalized on one of the two layers. The layers are then decoupled in agreement with the different predictions.\cite{Latil07, Dossantos07, Shallcross08, Bistritzer10, Bistritzer11, Bistritzer11_PNAS, Suarez10, Trambly10, Trambly12, Omid14}
Doping does not affect this result as shown 
Figure \ref{Fig_participaionRatio_dope}(a) for  $(5,9)$ bilayer.

For intermediate  $\theta$ values --bilayers $(6,7)$ ($\theta=5.08^o$) figure \ref{Fig_participaionRatio_dope}(b) 
and $(12,13)$ ($\theta=2.65^o$) figure \ref{Fig_participaionRatio_dope}(c)-- 
the participation ratio of a state slightly depends on the energy. It is closer to one in the energy range of the vHs where the interaction between the two planes is stronger and closer to 0.5 in the vicinity of the Dirac energy (interaction between layer is smaller for these energies). When the bilayer is doped, the energy region where the interaction between planes is weaker is just shifted accordingly. 

For very small $\theta$ values, 
--bilayer $(25,26)$ ($\theta=1.30^o$) figures \ref{Fig_participaionRatio} and \ref{Fig_participaionRatio_dope}(d) and $(33,34)$ ($\theta=0.99^o$)--
states with energy around $0$ are strongly localized (small $\langle p\rangle$ values).
An analysis of spacial repartition of eigenstates, 
shows that theses states are localized on the AA zones
of the moir\'e (see  Ref. \onlinecite{Trambly10, Trambly12}).  
For instance, the participation ratio of one eigenstate at Dirac point is $p\simeq0.12$
in $(33,34)$.\cite{Trambly10} The peak remains but shifted in the doped case (figure \ref{Fig_participaionRatio_dope}(d)).

We also define  a participation ratio per layer by
\be
p^*(\psi) = \frac{1}{2 \,  \left( P_1^{2} +  P_2^{2} \right) },
\ee
where $ P_l$, $l =$ 1 and 2, are the weight of the eigenstate $\psi$ on layer 1 and 2, 
respectively:
\be
P_l = {\sum_{i_l} |\langle {i_l} | \psi \rangle|^2 } \,
\ee
where $|i_l \rangle$ are the $p_z$ orbitals on the atoms $i_l$ of the layer $l$. 
An eigenstate with non zero weight only in one layer corresponds to $p^*=1/2$, 
whereas $p^*=1$ for an eigenstate uniformly 
delocalized on the two layers.
The average layer participation ratio $\langle p^*\rangle$ at energy $E$ is presented on 
figure \ref{Fig_participaionRatio}(b) for neutral bilayers and figure \ref{Fig_participaionRatio_dope}(a$^*$-d$^*$) for doped ones.
For large $\theta$ angles, states exist only in one of the two layers ($\langle p^*\rangle \simeq 1/2$) whatever the doping is 
(figures \ref{Fig_participaionRatio}(b) and  \ref{Fig_participaionRatio_dope}(a$^*$)).
As $\theta$ decreases, $\langle p^*\rangle$ increases, which shows that states spread more and more on the two layers. 
For very small $\theta$, states of undoped cases are uniformly distributed  on the two layers for all energies (figure \ref{Fig_participaionRatio}(b)).
Figure \ref{Fig_participaionRatio}(b) also shows that for intermediate angles, the distribution of eigenstate on the two layers is weaker close to the Dirac energy than for energy in the vicinity of the vHs.  
For intermediated $\theta$ and very small $\theta$, higher doping also seems to decrease the distribution on the two layers (figures \ref{Fig_participaionRatio_dope}(c$^*$) and \ref{Fig_participaionRatio_dope}(d$^*$)). 

While doping does not change qualitatively the average participation ratio $\langle p\rangle$, it significantly decreases  the average participation ratio per layer $\langle p^*\rangle$.
Therefore an asymmetric doping in a rotated bilayer favors a decoupling of the states between the layers.
For very small angles (figures \ref{Fig_participaionRatio_dope}(d) and \ref{Fig_participaionRatio_dope}(d*), 
$\theta = 1.30^{\rm o}$),
the localization in AA zone is obtained in doped like in undoped bilayer. At all energies around Dirac energy,
states are distributed on the two layers, $0.9 \le \langle  p^* \rangle  \le 1$, 
and the doping reduces a little bit 
the equal repartition of each eigenstate between the two layers in undoped bilayer, 
$\langle  p^* \rangle{\rm (doped)}  < \langle  p^* \rangle{\rm (undoped)} \simeq 1$.

\section{Quantum diffusion in bilayers}
\label{SecTransportBilayer}

\begin{figure}
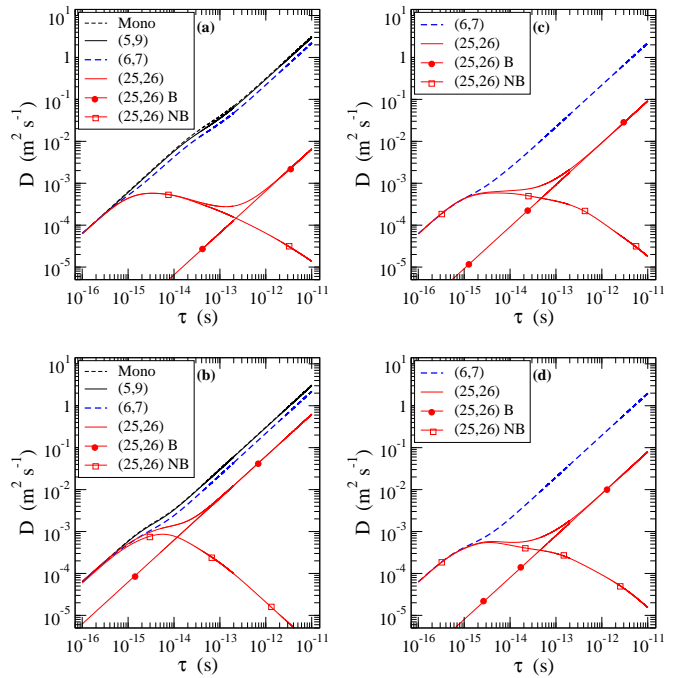


\includegraphics[width=4.2cm]{Fig_Transp_Dif_e0_dec14.eps}
~\includegraphics[width=4.2cm]{Fig_Transp_DEps.2_Dif_e-.1_dec14.eps}

\vskip 0.3cm

\includegraphics[width=4.2cm]{Fig_Transp_Dif_e.1_dec14.eps}
~\includegraphics[width=4.2cm]{Fig_Transp_DEps.6_Dif_e-.3_dec14.eps}

\caption{(color online) Diffusivity $D$ versus scattering time $\tau$
in graphene, (5,9), (6,7) and (25,26) bilayers.
For (25,26) bilayers the Boltzmann (B) term and non-Boltzmann (NB) term are shown.
$\Delta \epsilon=0$ for (a) energy $E_{\rm F}$ close to the Dirac energy ($E_D=0$), 
(b) $E_{\rm F}=0.1$\,eV. 
(c) $\Delta \epsilon=0.2$ and $E_{\rm F}=-0.1$\,eV
(d) $\Delta \epsilon=0.6$ and $E_{\rm F}=-0.3$\,eV
}
\label{Fig_Transp_Dif}
\end{figure}

In this section, we analyse the consequence on transport properties of the ``localization'' mechanism\cite{Bistritzer10,Bistritzer11,Bistritzer11_PNAS,Trambly10,Trambly12,Kim13_PRL} induced by the small rotation angles.
The conductivity along $x$-axis is given by the Einstein formula,
\be
\sigma_{xx} (E_{\rm F}) = \frac{e^2}{S} n(E_{\rm F}) \mathcal{D}(E_{\rm F}),
\ee 
where $n(E_{\rm F})$ and  $\mathcal{D}(E_{\rm F})$ 
are the total density of states per surface $S$ and the average diffusivity at the Fermi energy $E_{\rm F}$, respectively.
In the relaxation time approximation,\cite{Trambly06,Mayou08RevueTransp,Mayou00} 
the effect of disorder is taken into account by a scattering time $\tau$.
In this approach $\tau$ contains both elastic scattering times $\tau_e$ 
due to static defects (adatoms, vacancies...) 
and inelastic scattering time $\tau_i$ 
due to phonons or magnetic field (...): 
$\tau^{-1} = \tau_e^{-1} + \tau_i^{-1}$.
$\tau$ decreases when the temperature $T$ increases 
and when the concentration of static defects increases.
As explained in appendix \ref{SecMethodeQuantumTransp},
the diffusivity $\mathcal{D}$ can by determined at every energy $E$
as function of the scattering time $\tau$.
$\mathcal{D}$
is the sum of two terms, 
\be
\mathcal{D}(E_{\rm F},\tau) = \mathcal{D}_{\rm B}(\tau) + \mathcal{D}_{\rm NB}(E_{\rm F},\tau) , 
\label{Eq_Diffusivite_B_NB}
\ee
where $\mathcal{D}_{\rm B}(E_{\rm F},\tau) = V_B^2 \tau /3$ is the Boltzmann term and 
$\mathcal{D}_{\rm NB}$ the non-Boltzmann term. 
$\mathcal{D}_{\rm NB}$ comes from non-diagonal terms in the velocity operator 
(equation (\ref{Eq_DeltaX2_NB}) in the appendix). 
In crystals, it is related to inter-band transitions activated by 
elastic or inelastic scattering.
For large $\tau$,  $\mathcal{D}_{\rm NB}$ decreases when $\tau$ increases, 
and $\mathcal{D}_{\rm NB}\rightarrow 0$ 
when $\tau \rightarrow +\infty$. 
Thus in crystals, 
$\mathcal{D} \simeq \mathcal{D}_{\rm B}$ when $\tau \rightarrow +\infty$.

Diffusivity calculated for graphene and several bilayers is presented figure \ref{Fig_Transp_Dif} 
for different $E_{\rm F}$ values and for doped or undoped bilayers. 
For graphene and bilayers with large and intermediate rotation angles,  
$\mathcal{D} \simeq \mathcal{D}_{\rm B}$
at every energy. 
The only effect of non-Boltzmann term is a change in the slope of $\mathcal{D}(\tau)$ at scattering time 
$\tau \simeq \hbar/E $
as explained in appendix \ref{SecMethodeQuantumTransp}.
Eventually at small scattering time, $\tau \ll \hbar/E$, the inter-band transition between the two bands of each Dirac cone contribute significantly. In the case of graphene, with a first neighbor coupling hamiltinian, the non-Boltzmann term $\mathcal{D}_{\rm NB}$ is equal 
to Boltzmann term $\mathcal{D}_{\rm B}$ and then $\mathcal{D} = 2\mathcal{D}_{\rm B}$ for $\tau \ll \hbar/E$ (see equations (\ref{DifGrapheneD_B}) and (\ref{DifGrapheneD_NB}) in appendix \ref{SecMonoGraphene}). 
This effect is related with the phenomenon of jittery motion also called Zitterbewegung \cite{Castro09_RevModPhys} which is important in the optical conductivity. 
In graphene and bilayers with large rotated angles, it occurs for very small scattering time values to be significant experimentally.
But in rotated bilayers with very small rotation angle $\theta$, 
for states at energy where velocity is very small
(i.e. energies close to Dirac energy),  
the Boltzmann term in equation (\ref{Eq_Diffusivite_B_NB}) goes down and non-Boltzmann term 
becomes significant in the total diffusivity.
For instance, figure \ref{Fig_Transp_Dif}(a) shows that for 
(25,26) bilayers ($\theta = 1.3^{\rm o}$) at $E_{\rm F}=0$,
$\mathcal{D}$ is strongly affected by the non-Boltzmann term for realistic\cite{Wu07} 
$\tau$ values. It results in a smaller diffusivity with respect to graphene case, that is almost independant on scattering time for $\tau \simeq 10^{-14}$ -- $10^{-13}$s.
In asymmetric doped bilayer (figures \ref{Fig_Transp_Dif}(d) and \ref{Fig_Transp_Dif}(c)) similar effect occurs at energies with small Boltzmann velocity (figure \ref{Figure_VB_diag}).
This regime, called small velocity regime, where non-Boltzamnn terms dominate transport properties has already been observed in systems with complex atomic structure such as quasicrystals  and complex metallic alloys (see Refs \onlinecite{Trambly06,Triozon02,Trambly11,Trambly14,Trambly14_QC} and Refs in there).
Roughly speaking small velocity regime is reached when mean free path $l = V \tau$ of charge carriers is smaller than spacial extension $L_{wp}$
of the corresponding wave packet. In this case, semi-classical approximation breaks down and a pure quantum description is 
necessary to calculated transport properties. Twisted bilayer with very small rotation angle have a huge unit cell and a huge cell of the moir\'e, in which states at $E\simeq E_D$ are confined in AA zone \cite{Trambly12} and have then a very small velocity (figure \ref{Figure_VB_diag}). Typically the size of AA zone is $\sim 0.5P$, where $P$ is the moir\'e period (equation (\ref{EqPeriodeMoire})), and then the extension of confined states in AA zone is $L_{wp} \lesssim 0.5P$. 
As $P$ increases when $\theta$ increases, the condition $ V \tau < L_{wp} $ of the small velocity regime should be satisfied for $\theta$ small enough. 

In doped and undoped twisted bilayers, for energy which does not correspond to the a peak of localization in the DOS, the Boltzmann velocity is larger, and  the non-Boltzamnn effect is neglectable  (figure \ref{Fig_Transp_Dif}(b)).

\section{Conclusion}

To sum up, numerical calculations show that doped rotated layers with large rotation angle and reasonable doping (inducing a shift of the Dirac point smaller than 0.8 eV) still behave like decoupled layers as found experimentally.\cite{Cherkez15} 
This result is of particular importance for epitaxial graphene on the C face of SiC, at least for large rotation angles. In this case, the C plane closest to the interface dominates the transport 
because it is doped due to charge transfer from the interface. This charge transfer corresponds to a shift of the Dirac point of $0.2 - 0.3$\,eV above the Fermi level. Experiments  and theory \cite{Magaud09}  showed that it is decoupled from the substrate and thanks to a large  rotation angle stacking, it can also be decoupled from the other C planes. Here we show that actual asymmetric doping do not alter the layer decoupling so that this plane can exhibit isolated graphene like properties even if it is sandwiched between the interface and other C layers as observed 
experimentally.\cite{Sadowski06,Hass08_prl}

Thanks to the tight binding scheme, we have been able to address the important question of the effect of doping on rotated graphene bilayers with intermediate angle values corresponding to large cells of moir\'e. 
For a small symmetric doping, twisted layers with large and intermediate rotation angles keep their characteristics: linear band dispersions, renormalized band velocity at Dirac point (K point) and van Hove singularities, as expected experimentally.\cite{Cherkez15} 
But a large enough doping increases the renormalization of the velocity. For large angles, this new effect occurs for unphysical doping values, but for intermediated  angles, it occurs for accessible doping values, typically when $\Delta \epsilon = {\rm few}~ 0.1$\,eV.

For very small angles, electronic states remain confined in the AA region of the moir\'e whatever the doping is,
as in undoped bilayers.\cite{Trambly10,Trambly12}
Therefore, the regime of confinement by very large cells of the moir\'e is not destroyed by the doping, but localization energies are shifted with the doping rate.
In this later case, by conductivity calculations, show that the Bloch-Boltzmann model breaks down and strong interference quantum effects dominate transport properties.

\section*{Acknowledgments}
The authors wish to thank P. Mallet, J.-Y. Veuillen, V. Cherkez, C. Berger, W. A. de Heer
for fruitful discussions.
The numerical calculations have been performed at the Centre de Calculs (CDC),
Universit\'e de Cergy-Pontoise.
We thank Y. Costes and D. Domergue, CDC, for computing assistance. 
We acknowledge financial support from ANR-15-CE24-0017.

\appendix
\section{Quantum transport calculation}
\label{SecMethodeQuantumTransp}

\subsection{Average square spreading in perfect crystal at zero temperature}

In the framework of Kubo-Greenwood approach for calculation of the conductivity,
a central quantity is the average
quadratic spreading of wave packets of energy $E$ at time $t$
along the $x$ direction,\cite{Mayou88,Roche99,Mayou95,Mayou00,Trambly06,Mayou08RevueTransp,Trambly14_QC}
\begin{eqnarray}
  \Delta X^{2}(E,t) =\left\langle \Big(\hat{X}(t)-\hat{X}(0) \Big)^{2}\right\rangle_{E},
 \label{Def_1}
\end{eqnarray}
where $\hat{X}(t)$ it the Heisenberg representation of the position operator $\hat{X}$.
$\langle \hat{A} \rangle_{E}$ means an average of diagonal elements of the operator $\hat{A}$ over all 
states with energy $E$. 
The diffusivity at zero temperature, $\mathcal{D}(E)$, at energy $E$ is deduced from $\Delta X^{2}$,
\begin{eqnarray}
\mathcal{D}(E)  = \lim_{t \rightarrow + \infty} D(E,t),
 \label{Def_2}
\end{eqnarray}
with
\begin{eqnarray}
D(E,t) = \frac{\Delta X^{2}(E,t)}{t},
 \label{Def_2.2}
\end{eqnarray}
where $D(E,t)$ is called diffusion coefficient. 
In a 2-dimensionnal system with surface $S$, the DC-conductivity $\sigma_{xx}$ at zero temperature 
along the $x$-direction is given by Einstein formula:
\be
\sigma_{xx} (E_{\rm F}) = \frac{e^2}{S} n(E_{\rm F}) \mathcal{D}(E_{\rm F}),
\ee 
where $n(E)$ is the total density of states per $S$  and $E_{\rm F}$ the Fermi energy.

In pure crystals at zero temperature, once the band structure is calculated from the tight-binding Hamiltonian
the average quadratic spreading
can be computed exactly in the basis of Bloch states.\cite{Trambly06,Mayou08RevueTransp}
The average square spreading is the sum of two terms:\cite{Trambly06,Mayou08RevueTransp}
\begin{equation}
\Delta X^2(E,t) = V_{\B}^2 t^2 + \Delta X_{\NB}^2(E,t).
\label{Eq_DeltaX2}
\end{equation}
The first term is the ballistic (intra-band)  contribution
at the energy $E$.
$V_B$ is the Boltzmann velocity in the $x$ direction. 
The semi-classical theory
is equivalent to taking into account only this first term.
The second term (inter-band contributions),
$\Delta X^2_{\NB}(E,t)$,  is
a non-ballistic (non-Boltzmann) contribution.
It is
due to the non-diagonal
elements in the eigenstates basis $\{| n\rangle \}$  of the velocity operator $\hat{V}_x$, 
\begin{eqnarray}
\hat{V}_x = \frac{1}{i \hbar}~ \Big[ \hat{X} , \hat{H} \Big].
\end{eqnarray}
From the definition (\ref{Def_1}), one obtains,\cite{Mayou08RevueTransp}
\begin{widetext}
\begin{equation}
\Delta X_{\NB}^2(E,t) = 2 \hbar^2
\left\langle
\sum_{\vec k, n' (n' \ne n)}
\left[ 1 - \cos\left(\frac{(E_{\vec k,n}-E_{\vec k,n'})t}{\hbar} \right)\right] 
\frac{\left| \langle n\vec k | \hat{V}_x | n'\vec k \rangle \right|^2}{(E_{\vec k,n}-E_{\vec k,n'})^2}
\right\rangle_{E_{\vec k,n}=E} 
\label{Eq_DeltaX2_NB}
\end{equation}
\end{widetext}
where $E_{\vec k,n}$  is the energy of the eigenstate $| n \vec k \rangle$ computed by diagonalisation of the tight binding Hamiltonian in reciprocal space.
The average velocity --Boltzmann velocity-- along direction $x$ of the electrons at energy $E$ 
is obtained numerically from diagonal elements of $\hat{V}_x$,
\begin{equation}
V_B(E) = \sqrt{
\left\langle
\left| \langle n\vec k | \hat{V}_x | n\vec k \rangle \right|^2
\right\rangle_{E_{\vec k,n}=E} }. 
\label{Eq_calcul_VB}
\end{equation}

\subsection{Relaxation time approximation}

The effect of static disorder and/or decoherence mechanisms such as electron-electron scattering,
electron phonon interaction 
(temperature), is not considered in the above section.
This effect can be treated in a phenomenological way by introducing an inelastic scattering time $\tau$ 
in the relaxation time approximation (RTA).\cite{Trambly06}
$\tau$ may include elastic scattering time $\tau_e$  (due to static defects like vacancies or adatoms) 
and inelastic scattering time $\tau_i$ (due to phonon, electron-electron scattering, effect of magnetic field):
$\tau^{-1} = \tau_e^{-1} + \tau_i^{-1}$.
$\tau$ decreases when the temperature increases and/or static defects concentration  increases.
In actual graphene at room temperature, realistic values 
of $\tau_i$ are a few $10^{-13}$\,s.\cite{Wu07}
The conductivity can then be estimated by:
\be
\sigma_{xx} (E_{\rm F},\tau) = \frac{e^2}{S} n(E_{\rm F}) \mathcal{D}(E_{\rm F},\tau ),
\label{Eq_sigma_tau_i} 
\ee 
with diffusivity
\begin{eqnarray}
D(E_{\rm F},\tau)=\frac{1}{2\tau^2} \int_0^\infty \! \Delta X^{2}(E_{\rm F},t)\,{\rm e}^{-t/\tau} \, \mathrm{d}t \, ,
\label{EquationDfromtaui}
\end{eqnarray}
where $\Delta X^{2}(E,t )$ is the average square spreading in crystal without defects (equation (\ref{Def_1})).
Here the Fermi-Dirac distribution function is taken equal to its zero temperature
value. This is valid provided that the electronic properties vary smoothly on the
thermal energy scale $k_{\rm B}T$.
From equations (\ref{Eq_DeltaX2}) and (\ref{EquationDfromtaui}), $D$ is the sum of a Boltzmann contribution $D_{\rm B}$
and a non-Boltzmann contribution $D_{\rm NB}$:
\be
\mathcal{D}(E_{\rm F},\tau) = \mathcal{D}_{\rm B}(\tau) + \mathcal{D}_{\rm NB}(E_{\rm F},\tau) . 
\ee 

RTA has been used successfully to compute\cite{Trambly06}
conductivity in approximants of quasicrystals where quantum diffusion and localization effect play a essential role\cite{Trambly06,Trambly14_QC,Trambly14,Berger93,Belin93} 
and conductivity in organic semiconductors.\cite{Ciuchi11}
In this paper we show that quantum interferences have a also strong effect in transport properties of rotated bilayers with very small angles.

\subsection{Quantum transport in graphene}
\label{SecMonoGraphene}

In pure graphene, assuming a restriction of the Hamiltonian (equation (\ref{Eq_hamilt})) 
to the first neighbor interactions only, 
$\Delta X_{\NB}^2(E,t)$ is given by:
\begin{equation}
\Delta X_{\NB}^2(E,t) = \frac{V_B^2\hbar^2}{2 E^2} \left( 1 - 
\cos \frac{2E}{\hbar}t   \right) ,
\label{Eq_DeltaX2_NB_graphene}
\end{equation}
with
\begin{equation}
V_{_\B} = \frac{3 a \gamma_0}{2\sqrt{2} \hbar},
\end{equation}
where $\gamma_0$ is the coupling term between first neighbor $p_z$ orbitals.
At small time $t$, $t \ll \hbar/E$,  the Non-Boltzmann term 
is equal to Boltzmann 
term $\Delta X_{\NB}^2(E,t) \simeq V_{_\B}^2t^2$, 
thus $\Delta X^2(E,t) \simeq 2 V_{_\B}^2t^2$ and $D(E,t) \simeq 2V_{_\B}^2t$. 
Whereas for large $t$, the Boltzmann term dominates and  
$\Delta X^2(E,t) \simeq V_{_\B}^2t^2$ and $D(E,t) \simeq V_{_\B}^2t$.
The non-Boltzmann term is due to matrix elements of the velocity operator 
between the two bands 
(i.e. inter-band coupling between the hole and electron states having the same wavevector). 
These matrix elements imply that the velocity correlation function has also two parts: 
one constant and the other oscillating at a frequency $2 E/ \hbar $ where $E$ is the energy of the state. 
This is precisely the phenomenon of jittery motion also called Zitterbewegung. 
Note that in any crystal having several bands there are also components of 
the velocity correlation function which are oscillating at frequencies 
$(E_{\vec k,n}-E_{\vec k,n'})/ \hbar $. 
Therefore Zitterbewegung is quite common in condensed matter physics. 
For example approximants of quasicrystals present very strong Zitterbewegung effect and 
the non-Boltzmann contribution dominates the Boltzmann 
contribution.\cite{Trambly06,Triozon02,Trambly11,Trambly14}

With defects (static defects or phonons...) in RTA,
the diffusivity $\mathcal{D}(E_{\rm F},\tau)$
is also the sum of a Boltzmann term, 
\be
\mathcal{D}_{\rm B}(\tau) = \frac{1}{3} V_B^2 \tau ,
\label{DifGrapheneD_B}
\ee
and a non-Boltzmann term
\be
\mathcal{D}_{\rm NB}(E_{\rm F},\tau) = \frac{1}{3} V_B^2 \tau \left( \frac{\hbar^2}{\hbar^2 + 4 E_{\rm F}^2 \tau^2} \right).
\label{DifGrapheneD_NB}
\ee
For small scattering time, $\tau \ll \hbar/E$, the non-Boltzmann term $\mathcal{D}_{\rm NB}$ is equal 
to Boltzmann term $\mathcal{D}_{\rm B}$ 
and $\mathcal{D} = 2\mathcal{D}_{\rm B}$. 
For large $\tau$, $\tau \gg \hbar/E$, $\mathcal{D}_{\rm NB} \to 0$ 
and $\mathcal{D} = \mathcal{D}_{\rm B}$.
When $E_{\rm F}=0$ (i.e. Dirac energy), the non-Boltzmann term equals the Boltzmann term for all scattering times. 
On figures \ref{Fig_Transp_Dif}(a) and \ref{Fig_Transp_Dif}(b) this modification of $\mathcal{D}(\tau)$ at $\tau = \hbar/E$ is clearly seen.
But this limit case should be very difficult to obtain experimentally.
Similar results is obtained for twisted bilayers with large angle of rotation $\theta$, whereas for small $\theta$ this modification becomes larger showing that non-Boltzmann term are not neglectable anymore.

\end{document}